\documentclass[jamsb]{jaums}      
\usepackage{amsmath}
\usepackage{graphicx}

\theoremstyle{thmit} 

\theoremstyle{thmrm} 

\newtheorem*{oldproof}{Proof}

\title[Ambiguity in the Free Energy]{%
Ambiguity in the Determination of the Free Energy for a Model of the Circle Map}
\author{Brian G. Kenny}
\address{Department of Theoretical Physics,
    Research School of Physical Science and Engineering,
    Australian National University,
    Canberra ACT 0200, Australia.}
\author{Tony W. Dixon}
\address{Department of Mathematics \& Statistics,
     Curtin University of Technology,
     GPO Box U1987,
     WA 6845, Australia.}

\begin{document}

\maketitle

\begin{abstract}
We consider a simple model to describe the widths of the mode
locked intervals for the critical circle map.  Using two different
partitions of the rational numbers, based on Farey series and
Farey tree levels respectively, we calculate the free energy
analytically at selected points for each partition. It is found
that the result of the calculation depends on the method of
partition.  An implication of this is that the generalized
dimensions $D_q$ are different for each partition except when
$q=0$, i.e. only the Hausdorff dimension is the same in each case.
\end{abstract}



\section{Introduction}

On the basis of their numerical investigations, Jensen et
al.~\cite{JBB83,JBB84a} conjectured that for the critical circle
map the mode locked intervals form a complete Devil's staircase
whose complementary set is a Cantor set
with a fractal dimension $D \sim 0.87$.
Further, their numerical results indicated that this dimension
is universal for maps with a cubic inflection point.

Their results were obtained in a purely empirical fashion by
computing the width of smaller and smaller mode locked steps
and subtracting the summed widths from the total staircase length.

The set in question is the set of irrational windings, which
do not lead to mode locking. It is multifractal in nature and, more
precisely, is believed to have a universal Hausdorff dimension $D_H \sim 0.87$,
In order
to better characterize such sets, different thermodynamic
formalisms \cite{Halsey,Feig87} have been introduced and with
them, the idea of generalized dimensions $D_q$ \cite{HentProc}.

In this paper we point out that for the circle map, such a
description is not unique and depends upon the details of how the
rational numbers are partitioned. Although we find
the thermodynamics of the set for two distinct partitions is quite different,
the Hausdorff dimension {\emph is} the same in both cases.

\section{Model for Interval Widths}

In order to illustrate this ambiquity, we shall examine a simple model
for the widths of the mode locked intervals of the circle map and
show that two different partitions of the rationals lead to two
different functional forms for the free energy $F(\beta)$ (defined below) and
hence to different $D_q$. Suppose that the width $\Delta(P/Q)$ of
the mode locked interval for the critical circle map associated
with the winding number $P/Q$ is given by \cite{commentLanf}
\begin{equation}
\label{widths} \Delta(P/Q) = k Q^{-\delta},
\end{equation}
independent of $P$.  $P$ and $Q$ are coprime integers; $k$ and
$\delta$ are constants (we assume $\delta > 2$ ).

In this model we know the widths of the mode locked intervals,
however we do not know precisely where the left and right hand
ends of the interval are located on [0,1]. This means we do not
know exactly where the holes are located. However, when
experiments are performed on one-dimensional quasiperiodic systems
at the borderline of chaos \cite{BarkleyCumming}, it is only
possible to determine the `devil's staircase' for mode locked
intervals at rational winding numbers $P/Q$ where $Q$ is
relatively small. Nevertheless, it is possible to determine the
thermodynamics of the system including an estimate of the
Hausdorff dimension of the set of irrational windings.

In this paper, we use a simple model for these widths and are able
to construct the thermodynamics of this model. This leads to an
estimate of the Hausdorff dimension in agreement with that
obtained by more general considerations.

From Eq (\ref{widths}) we may analytically calculate the total sum
of the widths of the mode locked intervals (which is taken to be
one at criticality).
\begin{equation}
\label{totalsum}
{\displaystyle \sum_{Q}} {\displaystyle \sum_{P} \Delta(P/Q) }
                ={\displaystyle  k \sum_{Q=1}^{\infty} \phi(Q) Q^{-\delta}}
                = k {\displaystyle \frac{\zeta(\delta -1)}{\zeta(\delta )}}
                = 1  ,
\end{equation}
where $\zeta$ is the Riemann zeta function and $\phi(Q)$ is the
Euler function \cite{HardyWright}. The sum in Eq (\ref{totalsum})
is over all irreducible rationals $P/Q$, $P<Q$, and $\Delta(P/Q)$
is the width of the parameter interval for which the iterates of a
critical circle map lock onto a cycle of length $Q$, with winding
number $P/Q$. Therefore $k$ is fixed to be
\begin{equation}
\label{kvalue} k=  \frac{\zeta(\delta)}{\zeta(\delta -1)}~.
\end{equation}
Now the Hausdorff dimension is the largest value of $\beta$ for
which the sum
\begin{equation}
\label{partition}
\begin{array}{cl}
Z(\beta)  &= {\displaystyle \sum_{Q}}{\displaystyle  \sum_{P} {[\Delta(\frac{P}{Q})]}^{\beta}}\\
          &= k^{\beta} {\displaystyle \sum_{Q}  \phi(Q) Q^{-\delta \beta}}  \\
          &= k^{\beta} {\displaystyle \frac{\zeta(\delta \beta -1)}{\zeta(\delta \beta )}}
\end{array}
\end{equation}
diverges. It is clear from Eq (\ref{partition}) that there is a
pole singularity in the sum when $(\delta \beta -1) \rightarrow 1$
from above, i.e. $\beta \rightarrow 2/\delta$. We conclude
therefore that the Hausdorff dimension is given by
\begin{equation}
\label{betahvalue} \beta_H = 2/\delta .
\end{equation}

\section{Thermodynamic Formalism}

A method for extracting a spectrum of scalings from experimentally
or numerically generated strange (fractal) sets was introduced by
Halsey et al.~\cite{Halsey} using a ``thermodynamic formalism".
They found that the scaling properties of normalized distributions
lying upon such sets were characterized by two indices, namely
$\alpha$ which determines the strength of the singularities and
$f$ which describes how densely they are distributed. The spectrum
of singularities is described by giving the range of
$\alpha$-values and the function $f = f(\alpha)$.

Shortly thereafter Feigenbaum  \cite{Feig87} pointed out that the
work of Halsey et al.~was in fact the microcanonical version of a
thermodynamical formalism introduced by Vul, Sinai and Khanin
\cite{Vul} in 1984 some years earlier. See also Ruelle
\cite{Ruelle}. He elucidated the canonical version (canonical
paradigm = CP) in contrast to the microcanonical version
(microcanonical paradigm = MP).
In this formulation, the probability measure is assumed constant
on the set. This formalism introduces a free energy $F(\beta)$
which is related to the $f(\alpha)$ as indicated below. In the
following we use the notation of Feigenbaum \cite{Feig87}.

Consider a  dynamical system whose attractor can be hierarchically
represented as a set of $N_n$ intervals $I^{(n)}_i$, $i = 1\ldots
N_n$ of lengths $\Delta^{(n)}_i$ at the $n$-th level.  The free
energy is defined by
\begin{equation}
\label{freeenergydefn} {N_n }^{-F(\beta)}= \sum_{i} {|
\Delta^{(n)}_i |}^{\beta}
\end{equation}
where asymptotically $F$ becomes independent of $n$. The relation
between the MP quantities ( Halsey et al.~\cite{Halsey}) and the
CP quantities (Feigenbaum \cite{Feig87}) is given by
\begin{subequations}
\begin{eqnarray}
\alpha    &=& 1/F'(\beta) \label{conversion1} \\
f      &=& \beta -F(\beta)/F'(\beta) \label{conversion2} \\
q      &=&   -F(\beta) \label{conversion3} \\
\tau      &=& -\beta \label{conversion4} \\
D_q      &=& \beta /\left[1+F(\beta)\right] \label{conversion5}
\end{eqnarray}
\end{subequations}

\subsection{Free Energy for Farey Series Partition}

In order to proceed further, we first recall that for a given $Q$,
the number of irreducible rationals $P/Q$ is $\phi(Q)$. The Farey
series of order $Q$, $\{F_Q \}$, is a set containing the
monotonically increasing sequence of all irreducible rationals
$P'/Q'$ ($P'$ and $Q'$ are coprime) between 0 and 1 whose
denominator does not exceed $Q$ \cite{HardyWright}. Next we define
a sequence of integers $\{Q_n\}$ such that asymptotically
\begin{equation}
\label{Qnlimit} \frac{Q_{n+1}}{Q_n} \rightarrow \sqrt{2}.
\end{equation}
With these $Q_n$, we define a sequence of Farey series $\{
F_{Q_{n-1}} \}$, $\{ F_{Q_{n}} \}$, $\{ F_{Q_{n+1}} \}$ etc. It
follows that the number of rationals contained in $\{
F_{Q_{n+2}}\}$ {\em but not in} $\{ F_{Q_{n+1}}\}$, is
asymptotically twice the number contained in  $\{ F_{Q_{n+1}}\}$
but not in  $\{ F_{Q_{n}} \}$. We therefore have a simple
prescription for grouping the rationals into distinct exclusive
sets $\{\{ F_{Q_{n+1}}\}$ - $\{ F_{Q_{n}}\} \}$ whose membership
increases exponentially fast. We may re-express this statement by
saying that the ratio of the number of rationals contained in each
of these consecutive sets is given by
\begin{eqnarray*}
\frac{{\displaystyle \sum_{Q_{n+1}}^{Q_{n+2}} \phi(Q) }}
  {{\displaystyle \sum_{Q_{n}}^{Q_{n+1}} \phi(Q) }} &\sim&
\frac{{\displaystyle \int_{Q_{n+1}}^{Q_{n+2}} Q dQ}}
     {{\displaystyle \int_{Q_{n}}^{Q_{n+1}} Q dQ}}\\
&=& \frac{ {\displaystyle Q_{n+2}^2 - Q_{n+1}^2} }
{ {\displaystyle Q_{n+1}^2 - Q_{n}^2} } \\
&=& 2
\end{eqnarray*}
asymptotically.  We have used \cite{HardyWright}
\[
\sum_Q \phi(Q)\sim {\displaystyle  \frac{3}{\pi^2} Q^2}
\]
in the large $Q$ limit.

This construction defines a partition of the rationals from which
one may define the free energy $F(\beta)$ in this simple model
asymptotically through
\begin{eqnarray*}
2^{-F(\beta)}&=&
 \frac{{\displaystyle \sum_{Q_{n+1}}^{Q_{n+2}} \phi(Q) Q^{-\delta \beta} }}
  {{\displaystyle \sum_{Q_{n}}^{Q_{n+1}} \phi(Q)  Q^{-\delta \beta} }} \\
&\sim & \frac{{\displaystyle \int_{Q_{n+1}}^{Q_{n+2}} Q
~Q^{-\delta \beta}dQ}}
     {{\displaystyle \int_{Q_{n}}^{Q_{n+1}} Q ~Q^{-\delta \beta} dQ}} \\
&=& \frac{ {\displaystyle {[ Q^{2- \delta \beta}
]}_{Q_{n+1}}^{Q_{n+2}}  }  }
     { {\displaystyle {[ Q^{2- \delta \beta} ]}_{Q_{n}}^{Q_{n+1}}  }  } \\
&=& \frac{ {\displaystyle Q_{n+2}^{2- \delta \beta} - Q_{n+1}^{2-
\delta \beta}} }
{ {\displaystyle Q_{n+1}^{2- \delta \beta} - Q_{n}^{2- \delta \beta}} } \\
 &=&
(\sqrt{2})^{2- \delta \beta}
\end{eqnarray*}
from which we conclude that
\begin{equation}
\label{FEfromseries}
 F(\beta) = \frac{\delta \beta}{2} -1.
\end{equation}
The free energy has the same form as that for a single scale
Cantor set, which is not surprising since there only appears to be
a single length scale associated with this model. The Hausdorff
dimension is found from the condition
\[
 F(\beta = \beta_H) = 0
\]
i.e.
\[
\beta_H = \frac{2}{\delta}
\]
in agreement with the result of Eq (\ref{conversion4}). Using the
$f(\alpha)$ language
\[
 \alpha = f( \alpha) = \frac{2}{\delta}
\]
i.e. the $f(\alpha)$ curve collapses to a single point. It should
be noted that the expression for $F(\beta)$ quoted in Eq
(\ref{FEfromseries}) {\em in no way depends on the partition
defined through} Eq (\ref{Qnlimit}). Indeed we could choose {\em
any} $a>1$ such that
\[
\frac{Q_{n+1}}{Q_n} \sim \sqrt{a},
\]
which would lead instead to
\[
a^{-F(\beta)} \sim  (\sqrt{a})^{2- \delta \beta},
\]
and again we have
\[
F(\beta) = \frac{\delta \beta}{2} -1.
\]

A natural choice for the $Q_n$ that suggests itself is the
Fibonacci sequence, $Q_n = F_n$  in which case
\[
\frac{Q_{n+1}}{Q_n}  \sim  \frac{1}{\rho}
\]
where $\rho$ is the golden mean
\[
\rho = \frac{\sqrt{5} -1}{2}
\]
and
\[
a = \frac{ 1}{\rho^2} .
\]
In fact such a choice has been used \cite{ArtusoCvitKen} for
evaluating the free energy (or its equivalent) and the Hausdorff
dimension in the case of the real mode locking critical circle
map.  This is a different partition of the rationals which also
utilized Farey series.  The crucial point, however, in both
examples is that one may generate partitions of the rationals
whose membership increases exponentially fast from one partition
to the next; by construction the variation in size of the
denominator $Q$ in the set $\{ \{ F_{Q_{n+1}}\} - \{ F_{Q_{n}}\}
\}$ is {\em only of order} 1. By contrast the Farey level
construction which we shall discuss next allows enormous
variations in $Q$ at a given level. The smallest denominator is
$Q$ while the largest denominator is of order $(1/\rho)^Q$.

\subsection{Free Energy for Farey Tree Partition}

The Farey tree level partition of the rationals has been
frequently used in discussions of the circle map
\cite{ArtusoCvitKen,CvitSS} and one may readily calculate the free
energy using this scheme. Since the number of rationals doubles
from one level to the next, the free energy may be defined through
\begin{equation}
\label{FEdefntree}
\begin{array}{lll}
2^{-F(\beta)}&=&
 \frac{{\displaystyle \sum_{i} {[ Q_i^{(n+1)}]}^{-\delta \beta} }}
  {{\displaystyle \sum_{j} {[ Q_j^{(n)}]}^{-\delta \beta} }}
\end{array}
\end{equation}
when comparing level $n$ of the Farey tree with level $(n+1)$. The
$Q_j^{(n)}$   (resp. $Q_i^{(n+1)}$) are the denominators of the
rationals $P_j^{(n)} / Q_j^{(n)}$ (resp. $P_i^{(n+1)} /
Q_i^{(n+1)}$) at level $n$ (resp. $(n+1)$). In general, it is not
possible to evaluate Eq (\ref{FEdefntree}) analytically and solve
for $F(\beta)$. However it is possible to evaluate it {\em
exactly} at certain values of $\delta \beta$. From Eq
(\ref{FEdefntree}) we have
\begin{equation} \label{FEdefntree2}
2^{-F(\beta/\delta)} =
 \frac{{\displaystyle \sum_{i} {[ Q_i^{(n+1)}]}^{- \beta} }}
  {{\displaystyle \sum_{j} {[ Q_j^{(n)}]}^{- \beta} }}.
\end{equation}

\section{Comparison of Free Energies for Different Partitions}

Eq (\ref{FEdefntree2}) may be exactly evaluated for a number of values of $\beta$.
At $ \beta = 0$ we have
\begin{equation}
\label{FEtreecase1}
\begin{array}{cccll}
  \beta = 0: &~~ & 2^{-F(0)}~~~ &= {\displaystyle \frac{2^{n+1}}{2^n}}\\
&&&=2 &
\end{array}
\end{equation}
i.e.
\[
F(0) = -1.
\]
Here we have used the fact that the number of rationals at level
$n$ in the Farey tree is $2^n$. This result for $F(0)$ is {\emph in agreement}
with Eq (\ref{FEfromseries}) when $\beta$ is set = 0.

When $ \beta = -1$ we have
\[
\begin{array}{cccll}
  \beta = -1: &~~ &  2^{-F(-\frac{1}{\delta})} &=
  \frac{{\displaystyle \sum_{i} { Q_i^{(n+1)}} }}
       {{\displaystyle \sum_{j} { Q_j^{(n)}} }}  &\\
&&&= {\displaystyle \frac{3^{n+1}}{3^n}} &\\
&&&=3 &
\end{array}
\]
from which we conclude that
\begin{equation}
\label{FEvalue1} F(-\frac{1}{\delta}) = -\frac{\log 3}{\log 2}.
\end{equation}
Here we have used the fact that \cite{ArtusoCvitKen}
\[
{\displaystyle \sum_{j}  {Q_j}^{(n)} \sim 3^n }
\]
at the level $n$ of the Farey tree. However, Eq
(\ref{FEfromseries}) predicts that
\[
F(-\frac{1}{\delta}) = -\frac{ 3}{ 2}
\]
in disagreement with Eq (\ref{FEvalue1}). Therefore we see that we
get {\em different} values for the free energy at $\beta=-1$
according to the partition.

It is also possible to evaluate
$F(\beta)$ exactly at another point in the Farey tree construction
\[
\begin{array}{cccll}
  \beta = -3: &~~ &  2^{-F(-3/\delta)} &=
  \frac{{\displaystyle \sum_{i} {[ Q_i^{(n+1)}]}^3 }}
       {{\displaystyle \sum_{j} {[ Q_j^{(n)}]}^3 }}  &
\end{array}
\]
so
\begin{equation}
\label{FEvalue3}
F(-\frac{3}{\delta}) = -\frac{\log 7}{\log 2}.
\end{equation}
Here we have used the fact that \cite{ArtusoCvitKen}
\[
{\displaystyle \sum_{j}  {[{Q_j}^{(n)}]}^3 \sim  7^n }.
\]
The original partition predicts from Eq (\ref{FEfromseries}) that
\[
F(-\frac{3}{\delta}) = -\frac{5}{2}
\]
again in disagreement with Eq (\ref{FEvalue3}). These results are
summarized in Table \ref{tab1}.

\hspace{1cm}

\fbox{Table \ref{tab1} near here}

\hspace{1cm}

Therefore it is possible to {\em show exactly analytically} that a
different partition of the rationals yields {\em different} values
for the free energy at specific values of $\beta$ (and hence a
different singularity spectrum $f(\alpha)$) in this simple model
for the widths of mode locked intervals. From Eq
(\ref{conversion5}) we may express this in terms of the
generalized dimensions $D_q$. For the first partition
\[
\begin{array}{lll}
D_q &=& \beta /(1+ F(\beta))\\
&=& {\displaystyle \frac {2}{\delta}}
\end{array}
\]
independent of $q$.  Equivalently in terms of MP, Eq
(\ref{FEfromseries}) states
\[
\tau = \frac {2}{\delta} ~(q-1).
\]
Since \cite{Halsey}
\[
\tau = (q-1)~D_q,
\]
we have
\[
D_q = \frac {2}{\delta}
\]
once more.

We see that $D_q$ is no longer constant. Cvitanovi\'{c}
\cite{Cvitcomm} has found the $q(\tau) (=-F(\beta)$) curve for the
case $\delta=2$ using the Farey tree construction together with
accelerated convergence numerical techniques. It is clear that in
this case $q(\tau)$ is no longer linear in $\tau$ suggesting that
there is more than one length scale which at first sight seems
surprising in this simple model. Since $q(\tau)$ is no longer
linear in $\tau$ (as it was in the Farey series partition), $D_q$
is no longer a constant but varies with $q$.

\hspace{1cm}

\begin{figure}[ht] \centering
\includegraphics[width=0.8\textwidth]{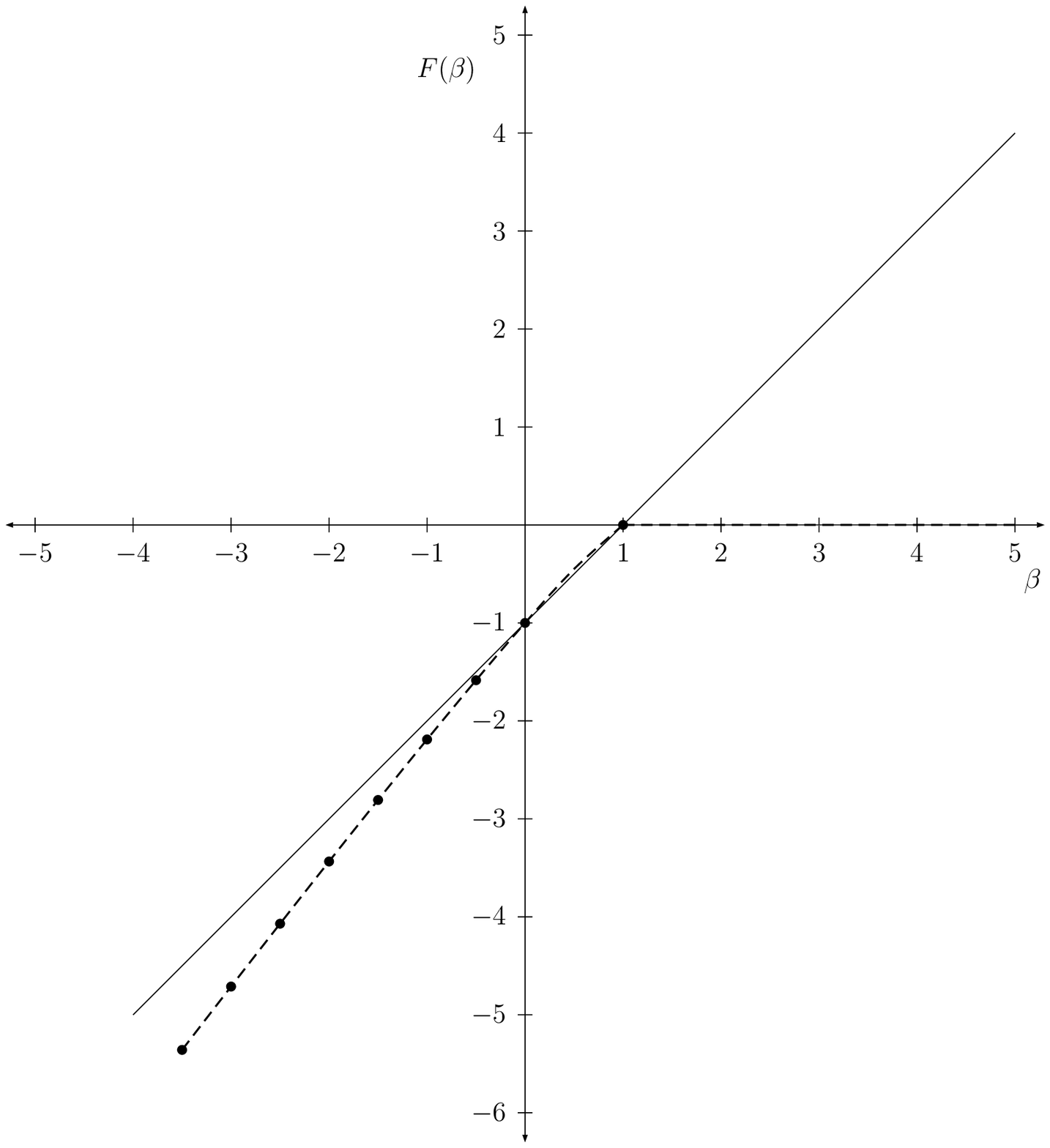}
\caption{ A comparison of the two partitions used to determine
$F(\beta)$. The continuous line is $F(\beta) = (\beta - 1)$ and is
associated with the Farey Series/Sequence partition. The broken
line shows $F(\beta)$ for the Farey tree partition. This maybe be
calculated analytically for negative half integer $\beta$ (see
Table \ref{tab1}) and  computationally when beta is positive for
finite order Farey tree levels. See \cite{ArtusoCvitKen} who
conjecture that asymptotically, $F(\beta)$ converges to $\beta =
1$ so that $F(\beta)$ for the Farey tree partition is $C^1$ at
$\beta = 1$. These different partitions lead to different
thermodynamic free energy $F(\beta)$ which have only two points in
common. One is at $F(0)= -1$ where all partitions must agree. The
other is at $F(1) = 0$ which means both partitions lead to the
same Hausdorff dimension $D_H=D_0 = 1$. Note that for the purposes
of illustration, the value $\delta=2$ has been chosen.}
\label{Figure1}
\end{figure}

\fbox{Figure \ref{Figure1} near here}

\hspace{1cm}

In Figure \ref{Figure1}, we plot $F(\beta)$ over a range of $\beta$
values for both partitions of the rationals. The straight line
$F(\beta)=(\beta-1)$ (shown as broken) is associated with the Farey series
partition. The other line (continuous) shows $F(\beta)$ computed
using Farey tree levels. For the purpose of illustration, the
value $\delta=2$ has been chosen. For negative half integer
$\beta$, $F(\beta)$ is known analytically for the latter partition
\cite{ArtusoCvitKen}. See Table \ref{tab2}. For positive $\beta$,
the thermodynamic function is computed for Farey tree levels 13,
15, 17. Using accelerated convergence techniques one then obtains
$F(\beta)=0$ for $\beta>1$ \cite{ArtusoCvitKen,Cvitcomm}. Clearly
the thermodynamic functions are quite different and this
conclusively demonstrates that their determination depends
strongly on the method used to partition the rationals.

\hspace{1cm}

\fbox{Table \ref{tab2} near here}

\hspace{1cm}

\section{Conclusion}

In conclusion we see that by analyzing an oversimplified model for
the widths of the mode locked intervals in a critical circle map
and partitioning the rationals $P/Q$  in two distinct ways, it is
possible to derive two completely different functional forms for
$F(\beta)$ [resp. $q(\tau)$] which only agree at two different
values of $\beta$ [resp. $\tau$]. As noted one of these is
$\beta=0$ (which was shown analytically in Eq (\ref{FEtreecase1}))
while the other is presumably at $-\tau=\beta=\beta_H$.
Cvitanovi\'{c} \cite{Cvitcomm} has shown that the latter is true
for $\delta=2$ to a high precision using numerical techniques.
Even if one has doubts about such numerical techniques, the
ability to compute the thermodynamic function analytically at
isolated points and find that the result depends on the form of
the partition used demonstrates unequivocally that a serious
ambiguity has arisen. One would expect that other partitions
different from those discussed here would yield thermodynamic
functions that are different from the ones depicted in Figure
\ref{Figure1}.

An alternative way of stating the conclusion is that the Farey
series scheme for partitioning the rationals when one adopts this
simplified model for the mode locking widths yields a {\em single
constant generalized dimension} $D_q=D_0$, while the Farey level
tree partition clearly yields a result for $D_q$ which is not
constant with variable $q$. Thus there is clearly ambiguity in
this attempt to determine the generalized dimensions associated
with the circle map which presumably persists for the real
critical circle map which is genuinely multifractal.

\section*{Acknowledgements}

BGK would like to thank Leo Kadanoff for his hospitality at the
James Franck Institute where part of this work was carried out as
well as for useful discussions and encouragement.


\newpage

\clearpage

\begin{table}
\centering
\begin{tabular}{|c|c|c|}
\hline
$\beta$      &  Farey series  &    Farey tree   \\
\hline
$1$     & 0              &        0                  \\
$0$     & -1             &        -1                  \\
$-1/2$  & $ -3/2 $   & $-\log 3/\log 2$  \\
$-3/2$  & $ -5/2 $   & $-\log 7/\log 2$\\
\hline
\end{tabular}
\caption{Thermodynamic free energy $F(\beta)$ evaluated at
selected values of $\beta$ for both partitions with $\delta=2$.
\label{tab1}  }
\end{table}

\clearpage

\begin{table}
\centering
  \begin{tabular}{|c|c|}
\hline
$\beta$  &   Asymptotic limit of $2^{- F(\beta)}$ \\
\hline
1        &   1\\
0        &   2\\
-1/2      &   3\\
-1        &   ~$(5 + \sqrt{17})/2$\\
-3/2      & 7\\
-2        &  ~$(11 + \sqrt{113})/2$\\
-5/2      &  ~$7 + 4\sqrt{6} $\\
-3        & $~~~~26.20249\ldots$\\
-7/2      & $~~~~41.0183\ldots$\\
\hline
\end{tabular}
\caption{Analytic limits of the ratio
${\displaystyle \sum_{i} {[Q_i^{(n+1)}]}^{-2\beta} /  \sum_{j} {[Q_j^{(n)}]}^{-2\beta} }$
for consecutive levels in the Farey tree in the asymptotic limit
$n\rightarrow \infty$ for various $\beta$ (see Eq (\ref{FEdefntree})).
\label{tab2}  }
\end{table}
\end{document}